# Geant4 Assisted Monte Carlo Prediction of Space Proton-Induced Soft Error Rate Based on Heavy Ion Testing Data

Artur M. Galimov, Regina M. Galimova, Gennady I. Zebrev

*Abstract* —A procedure of space proton-induced Soft Error Rate calculation based on heavy ion testing data is proposed. The approach relies on Geant4-assisted Monte Carlo simulation of the secondary particle LET spectra produced by proton interactions.

*Index Terms*— Cross section, LET spectrum, modeling, protons, single event upsets, soft error rate

## I. INTRODUCTION

High susceptibility of the integrated circuits (IC) to the proton-induced soft errors remains an actual problem of the device reliability. It is well-known that high energy space protons do not produce single event upsets in the IC via direct ionization because of low Linear Energy Transfer (LET). However, the proton-induced upsets can occur due to the ionization by the secondary products of proton-induced nuclear interactions. The dominant abundance of energetic protons in space, especially in South Atlantic Anomaly (SAA) region, makes them the main contributor to soft error rate (SER) [1].

The situation with the proton-induced upsets gets worse in the modern commercial off-the-shelf (COTS) devices. Along with the critical charge reduction in these devices, the presence of the high-Z materials in the back-end-of-line (BEOL) increases the IC sensitivity. The LETs of the secondary particles from the 500 MeV proton-W collisions may reach 34 MeV-cm$^2$/mg [2]. The particles with such high LET values can produce not only the single-bit but also the multiple cell upsets in the modern COTS devices [3]. The processes of nuclear interactions are very complicated and generally required Monte Carlo [4] and TCAD numerical calculation. Several Monte Carlo Geant4 based tools (MRED [5], MUSCA SEP [6] etc.) have been developed to simulate the processes at each level of radiation response of the ICs: transport of radiation, charge deposition and collection, the transistor and circuit response. These tools contain TCAD simulators which require detailed information about internal structure of the ICs that makes it difficult to use them in practice. We have proposed here a simplified Monte Carlo approach to the proton-induced SER problem based on the data for heavy ion direct ionization. Using the Geant4 tool, we simulate the LET spectra of secondary particles in the single thin lamina IC sensitive volume followed by the calculation of SER in the same way as for heavy ions [7]. This report contains preliminary validation of this simplified approach.

## II. SIMPLIFIED APPROACH

We consider here the LET spectra of the primary heavy ion $\phi_{HIV}(\Lambda)$ and the secondary products of nuclear interactions $\phi_{SEC}(\Lambda)$ on equal grounds. Then, the total SER can be calculated as follows

$$R_{total} = R_{HIV} + R_P = \int_0^\infty \sigma(\Lambda)\left[\phi_{HIV}(\Lambda) + \phi_{SEC}(\Lambda)\right] d\Lambda, \quad (1)$$

where $\sigma(\Lambda)$ is the ion-induced SEU cross-section averaged over all directions, $\phi_{HIV}(\Lambda)$ is the omnidirectional differential LET spectrum of incident particle, $\phi_{SEC}(\Lambda)$ is the differential LET spectrum of secondary particles from the proton-nucleus collisions. The first term of (1) is calculated commonly, generating the LET spectra for ions excluding protons, e.g. in the OMERE tool [8]. The secondary particle spectrum $\phi_{SEC}(\Lambda)$ is simulated in Geant4 using the energy spectrum of protons on a given orbit. Based on Geant4 v.10.3 library the custom simulator of proton-material interaction has been developed. The Geant4 General Particle Source (GPS) is used to model the radiation environment. The virtual target represents a multi-layered planar structure. Note that in the frame of compact modeling [7] we use the concept of continuous thin lamina as a sensitive volume (SV) of the device. We simulate the ionization from the products of nuclear reactions within a single thin detector with 50 nm thickness. The width and length of the detector are chosen of order of the circuit size (1 cm). This allows us to reduce the fluence of projectile protons to $10^7$ cm$^{-2}$ without a using of the biasing technique [5] in contrast to traditional approach [9, 10]. For benchmarking purposes, the LET spectrum of secondary particles $\phi_{SEC}(\Lambda)$ was compared with the Hiemstra results [11].

A. M Galimov, G. I. Zebrev are with Department of Micro- and Nanoelectronics of National Research Nuclear University MEPHI, 115409, Kashirskoe sh., 31, Moscow, Russia, e-mail: farmat913144@gmail.com

R. M. Galimova is with Institute of Computational Mathematics and Information Technologies of Kazan Federal University KFU, 420000, Kremlevskaya st., 18, Kazan, Russia.

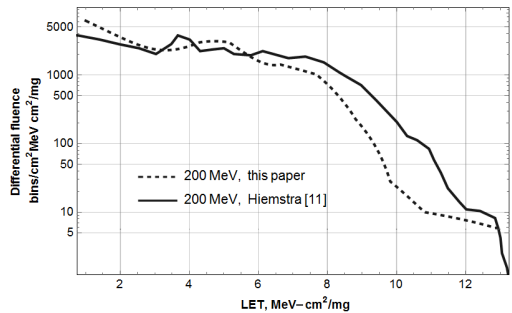

Fig. 1. Comparison of our Monte Carlo calculation with the Hiemstra simulation for interactions of 200 MeV protons with the silicon target. Both spectra are normalized to $10^{10}$ protons.

As can be seen in Fig. 1 the both spectra are rather close.

### III. MODEL VALIDATION

The proposed approach has been validated by comparison with in-flight data that has already carried out for heavy ions [7]. The energy spectra of protons in given orbits behind the shielding were generated in the OMERE tool. The Geant4 GPS differential histogram mode was used to load the generated energy spectrum of protons into simulation environment. We assume for the sake of simplicity that the on-board SRAMs don't consist high-Z materials in the BEOL. Fig. 2 shows the simplified overlayer planar structure used in all simulation sets.

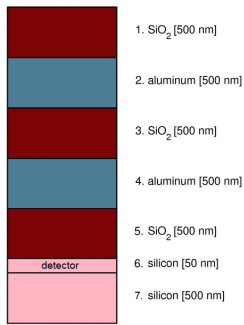

Fig. 2. The view of the simplified stack of devices under the model. The picture obtained in the CRÈME-MC website.

We obtained the spectrum of secondary particles produced in the given proton fields. Fig. 3 shows simulated and actual heavy ions spectra in orbits of Proba-2 and SAC-C satellites. As can be seen from Fig.3, the spectra of secondary particles have a much harsh part at high LETs than cosmic heavy ions spectra.

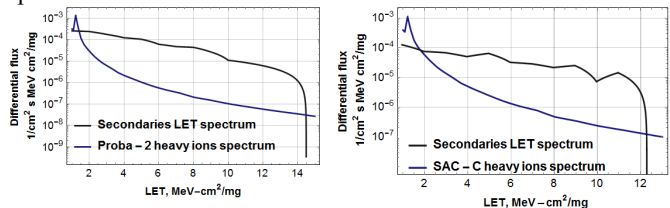

Fig. 3. Simulated secondary particles and cosmic heavy ions differential LET spectra.

The simulated spectra $\phi_{SEC}(\Lambda)$ were substituted in (1) to calculate the total on-orbit SERs. Table I shows a comparison of in-flight and the calculated SERs for heavy ions and protons of SRAMs with the known cross-section data [1, 12].

TABLE I. SER COMPARISON COMPENDIUM

| Memory circuit | Heavy ions SER (upsets/ day) | | Protons SER (upsets/ day) | |
|---|---|---|---|---|
| | In-flight | Compact Model Prediction | In-flight | Compact Model Prediction |
| AT68166 | 0.168 | 0.113 | 1.75 | 2.73 |
| HM628512 | 0.05 | 0.07 | 1.03 | 1.40 |
| KM6840003 | 0.35 | 0.73 | 3.59 | 3.49 |
| IS62W20488 | 0.386 | 0.1 | 2.79 | 2.85 |

Calculation results of on-orbit SERs for several SRAMs show a good correlation with the in-flight data.

### CONCLUSION

We have been proposed a simplified procedure to proton-induced SER calculation relying upon the heavy ions cross-section data. The physical basis for this approach is that in both cases the cause of SEU is direct ionization from the primary or the secondary heavy ions. The proposed Geant4 assisted approach was validated by direct comparison with the on-board results.

### REFERENCES

[1] C. Boatella, G. Hubert, R. Ecoffet, and F. Bezerra, "ICARE on-board SAC-C: More than 8 years of SEU & MBU, analysis and prediction," *IEEE Trans. Nucl. Sci.*, Vol. 57, No. 4, pp. 2000-2009, Aug. 2010.
[2] J. R. Schwank, M. R. Shaneyfelt, J. Baggio, P. E. Dodd, J. A. Felix, V. Ferlet-Cavrois, P. Paillet, D. Lambert, F. W. Sexton, G. L. Hash, and E. W. Blackmore, "Effects of particle energy on proton-induced single event latchup," IEEE Trans. Nucl. Sci., vol. 52, no. 6, pp. 2622 – 2629, Dec. 2005.
[3] D. Giot, P. Roche, G. Gasiot, J.-L. Autran, and R. Harboe-Sørensen, "Heavy Ion Testing and 3-D Simulations of Multiple Cell Upset in 65 nm Standard SRAMs," *IEEE Trans. on Nucl. Sci.*, vol. 55, no. 4, pp. 2048-2054, Aug. 2008.
[4] S. Agostinelli *et al.*, "Geant4—A simulation toolkit," *Nucl. Inst. Meth. Phys. Res. A—Accelerators, Spectrometers, Detectors and Associated Equipment*, vol. 506, no. 3, pp. 250–303, 2003.
[5] R. A. Weller, M. H. Mendenhall, R. A. Reed, R. D. Schrimpf, K. M. Warren, B. D. Sierawski, and L. W. Massengill, "Monte Carlo simulation of single event effects," *IEEE Trans. Nucl. Sci.*, vol. 57, no. 4, pp. 1726–1746, Aug. 2010.
[6] G. Hubert, S. Duzellier, C. Inguimbert, C. Boatella-Polo, F. Bezerra, and R. Ecoffet, "Operational SER Calculations on the SAC-C Orbit Using the Multi-Scales Single Event Phenomena Predictive Platform (MUSCA-SEP3)," *IEEE Tran. Nucl. Sc*i., vol. 56, pp. 3032–3042, Dec. 2009.
[7] G. I. Zebrev, A. M. Galimov, "Compact Modeling and Simulation of Heavy Ion Induced Soft Error Rate in Space Environment: Principles and Validation," to be published in *IEEE Trans. Nucl. Sci.*, Vol. 64, No. 4, Aug. 2017, available at ieeexplore.ieee.org.
[8] Outil de Modélisation de l'Environnement Radiatif Externe OMERE [Online]. Available: http://www.trad.fr/OMERE-Software.html
[9] J. H. Adams *et al.*, "CRÈME: The 2011 revision of the cosmic ray effects on micro-electronics code," *IEEE Trans. Nucl. Sci.*, vol. 59, no. 6, pp. 3141–3147, Dec. 2012.
[10] R. Ladbury, J.-M. Lauenstein and K. P. Hayes, "Use of Proton SEE Data as a Proxy for Bounding Heavy-Ion SEE Susceptibility," *IEEE Trans. Nucl. Sci.*, Vol. 62, No. 6, p. 2505 – 2510, 2015
[11] D. M. Hiemstra and E. W. Blackmore, "LET Spectra of proton energy levels from 50 to 500 MeV and their effectiveness for single event effects characterization of microelectronics," *IEEE Trans. Nucl. Sci.*, vol. 50, no. 6, pp. 2245–2250, Dec. 2003.
[12] M. D'Alessio, C. Poivey, V. Ferlet-Cavrois, P. Matthijs, "SRAMs SEL and SEU In-flight Data from PROBA-II Spacecraft," *RADECS-2013 Proceedings*, 2013.